# An Internet-based Audience Response System for the Improvement of Teaching

Rainer Lütticke, Bochum University of Applied Sciences, Institute of Computer Science
rainer.luetticke@hs-bochum.de
Ridvan Cinar, Bochum University of Applied Sciences, Institute of Computer Science

**Abstract:** We have developed an audience response system using the Internet. In this way we combine the advantages of common audience response systems using handheld devices and the easy and cheap access to the Internet. Therefore our system can be used in many different teaching/learning processes which take place in rooms where Internet access is given, e.g. in computer rooms of universities or schools, or in rooms with wireless Internet access using laptops.

## Common Audience Response Systems

An audience response system allows a group of people to answer a question or to vote on a topic. Each person in a room has a remote control with which selections can be made. Selections are communicated to a computer via receivers in this room. After a set time is over, the system permits further answering and tabulates the results. Normally, the results are immediately made available to the participants via graphics displayed on projector.

Most audience response systems use a combination of software and hardware to present questions, record responses, and provide feedback. The hardware consists of two components: the receiver and the remote controls. These handheld devices have several labelled buttons corresponding to the answers to a question posed by the lecturer.

Questions are created using standard software or special audience response system software. They are displayed on a screen and the audience responds by entering their answers using the devices.

There exist many commercial audience response systems on the market using such interactive handheld devices (e.g. Burnstein & Lederman, 2003). This fact and educational studies (e.g. Uhari, Renko, & Soini, 2003; Latessa & Mouw, 2005; Plischko, 2006) show that teaching-learning scenarios including audience response systems are successful. That means on one side students have learned more by this kind of interaction and on the other side teachers can fit his teaching style or lecture content on the answers of the audience. Encouraging participation, increasing retention, and adding excitement of the students as well as immediate feedback to answers to the whole group for evaluational purposes are the advantages of audience response systems.

However, commercial systems are relatively expensive so that their use in universities and schools is currently rare. Another disadvantage is that the number of students in such a teaching-learning scenario is limited to the number of bought handheld devices. A problem which is especially hard in large courses at universities. Additionally, commercial systems are not or not enough expandable to individual requests (i.e. new functions) of authors. In case of technical problems servicing an be executed only by experts.

## Our Audience Response System

Solving the mentioned problems we have developed an audience response system in which the students don't interact to answers of the teacher via handheld devices but via PC or laptop connected with the Internet. In computer rooms the access to the Internet may be given by wire. Using laptops a wireless access must be given. Especially in universities, where more and more students have a laptop and wireless Internet access is given, our system can be used. Interaction via Internet has the advantage that the number of students using the audience response system is not an important parameter for the system.

Furthermore, our system can be used in teaching-learning scenarios where the students and teacher are not in the same room (e.g. live video streaming).

If no learning management system is established the teacher has also the possibility to use our system for online exercises as homework after the lecture.

### Design of the System

In our system not only the students interact via Internet. Also the teachers construct their questions via Internet and use it to evaluate of the answers to the questions.

The installation of the system requires no particular effort. The system is constructed (end)user-friendly and comprehensible, so that there is no need for initial training. There exists an easy authoring tool with which the teacher can create single choice and multiple choice questions and answers. The number of answers is not limited and several questions can combined to a question group.

To create a new question group the teacher can also use his already existing questions (question pool). The same questions can be deployed in different groups without any effect on the statistics. Each question group can be individual locked, unlocked or edited. Such a question group (which can also consist only of one question) is put on a web page which is linked by the start page of our audience response system.

During a lecture a teacher can ask his students to select a destined question group and to answer the question(s). After the time for question answering is over the given answers are written in a database. Immediately the teacher can check the answers which are statistically presented in web page of our system. The fraction of answers is visualized by colored statistical bars.

Since all answers with additional information about group and time are saved in a database statistics between different groups and different times can be made.

Of course, web pages of our authoring tool, the statistics of answers, and the start page for the students are password protected. Although our system is primarily designed for educational purposes, it is also possible to create polls with a public participation without any access password for example like opinion or market surveys.

The web pages are dynamically generated using PHP and the data are saved in and read from a MySQL database. The user needs only a web browser and the URL of our system to answer questions. Therefore there is no software setup needed on the user site and minimal computing power is used.

Actually we are testing our system. Next summer term we will use it in regular teaching courses at our university of applied sciences. We expect large interest of the students on the system because the reactions on tests were very positive.